\documentclass[12pt,preprint]{emulateapj}

\usepackage{longtable}
\usepackage{color}
\usepackage{rotating}
\usepackage{natbib}
\usepackage[pdftitle={Novel Method for Identifying Exoplanetary Rings},
  colorlinks = true, breaklinks = true, citecolor = blue, linkcolor =
  blue, urlcolor = blue, pdfauthor = {Jorge Zuluaga et al.}]{hyperref}
\newcommand{\beq}[1]{\begin{equation}\label{#1}}
\newcommand{\eeq}{\end{equation}}
\newcommand{\sub}[1]{_{\rm #1}}
\newcommand{\beqn}{\begin{eqnarray}}
\newcommand{\eeqn}{\end{eqnarray}}

\newcommand{\Rs}{R_\star}
\newcommand{\Rp}{R\sub{p}}

\newcommand{\iR}{i\sub{R}}
\newcommand{\iP}{i\sub{R}}

\newcommand{\dg}{^{\circ}}
\newcommand{\Rext}{R\sub{e}}
\newcommand{\Rint}{R\sub{i}}
\newcommand{\fext}{f\sub{e}}
\newcommand{\fint}{f\sub{i}}
\newcommand{\tT}{T\sub{14}}
\newcommand{\tF}{T\sub{23}}
\newcommand{\ARp}{A\sub{Rp}}

\newcommand{\Rpobs}{p\sub{obs}}
\newcommand{\aobs}{a\sub{obs}}

\newcommand{\rhoobs}{\rho\sub{\star,obs}}
\newcommand{\rhotrue}{\rho\sub{\star,true}}
\newcommand{\PR}{{\rm PR}}
\newcommand{\reff}{r}
\newcommand{\hl}{}
\newcommand{\mhl}{}

\renewcommand{\ss}{\scriptstyle}

\begin{document}

\title{A novel method for identifying exoplanetary rings}

\author{Jorge I. Zuluaga \altaffilmark{1,2,3}, David
  M. Kipping\altaffilmark{1,4}, Mario Sucerquia\altaffilmark{2} and
  Jaime A. Alvarado\altaffilmark{2}}

\altaffiltext{1}{Harvard-Smithsonian Center for Astrophysics,
  Cambridge, MA 02138, USA}

\altaffiltext{2}{FACom-Instituto de F\'{\i}sica-FCEN, Universidad
  de Antioquia, Calle 70 No. 52-21, Medell\'{\i}n, Colombia}

\altaffiltext{3}{Fulbright Visitor Scholar}

\altaffiltext{4}{Menzel Fellow}
		 
\begin{abstract}

The discovery of rings around extrasolar planets (``exorings'') is one
of the next breakthroughs in exoplanetary research. Previous studies
have explored the feasibility of detecting exorings with present and
future photometric sensitivities by seeking anomalous deviations in
the residuals of a standard transit light curve fit, at the level of
$\simeq100$ ppm for Kronian rings. In this work, we explore two much
larger observational consequences of exorings: (1) the significant
increase in transit depth that may lead to the misclassification of
ringed planetary candidates as false-positives and/or the
underestimation of planetary density; and (2) the so-called
``photo-ring'' effect, a new asterodensity profiling effect, revealed
by a comparison of the light curve derived stellar density to that
measured with independent methods (e.g., asteroseismology). While
these methods do not provide an unambiguous detection of exorings, we
show that the large amplitude of these effects combined with their
relatively simple analytic description, makes them highly suited to
large-scale surveys to identify candidate ringed planets worthy of
more detailed investigation. Moreover, these methods lend themselves
to ensemble analyses seeking to uncover evidence of a population of
ringed planets. We describe the method in detail, develop the basic
underlying formalism and test it in the parameter space of rings and
transit configuration. We discuss the prospects of using this method
for the first systematic search of exoplanetary rings in the Kepler
database and provide a basic computational code for implementing it.

\end{abstract}		 


\maketitle  

\section{Introduction}
\label{sec:introduction}

Since the discovery of the first transiting planet
\citep{Charbonneau00,Henry00}, planetary eclipses have emerged as
powerful tools for characterizing exoplanets. Numerous novel methods
have been devised using transits to identify non-conventional
planetary properties, such as oblateness \citep{Carter10,Leconte11},
magnetic bow shocks \citep{Vidotto10}, exomoons
\citep{Kipping09,Heller14}, and exoplanetary rings or ``exorings''
\citep{Barnes04,Ohta09,Tusnski11,Mamajek12,Kenworthy15}.

With the advent of new instruments surveying the sky for transits with
precise photometry (PLATO, \citealt{Rauer2011}; EELT,
\citealt{Guyon2012}; GMT, \citealt{Johns2012}; TESS,
\citealt{2014RickerTESS} and JWST, \citealt{Beichman2014}), there is
great potential for discoverying new and unconventional exoplanetary
phenomena in the coming decade.

The discovery of exorings would be particularly interesting.  All of
the solar system's giant planets have rings and at least one, Saturn,
has sufficiently extended ring systems that are sufficiently extended
to produce observable signatures with current/future instrumentation
\citep{Barnes04,Ohta09}.  The discovery and characterizarion of
exorings could shed light on important planetary processes, such as
planetary and moon formation \citep{Mamajek12} and planetary interior
structure \citep{Schlichting11}.

\citet{Barnes04} developed the first theoretical model of exoring
transits.  They showed that, provided photometric sensitivities and
time resolution of $1-3\times 10^{-4}$ and $\sim$ 15 minutes, light
curve residual analysis could be used to resolve and characterize
rings. More recently, \citet{Ohta09} showed that the presence of rings
could produce spectroscopic signatures, detectable by ground-based
telescopes, for sensitivities $<0.1$ m $s^{-1}$. Independently,
\citet{Tusnski11} have tested a model of ring transits showing that
the properties of hypothetical rings can be reliably recovered.

Despite these theoretical advances, there have been no systematic
surveys for exorings with archival photometry. One reason for this is
the assumption that a practical ring survey would employ the same
technique as that for their discovery and characterization, namely,
the detailed analysis of light curves via fits to complex ring transit
models.

Here, we present a novel method for perfoming systematic searches for
exoring candidates.  Our method relies only on the measurement of the
basic transit parameters (i.e. depth, duration), avoiding the need for
computationally expensive fits of a transit ring model.  The method
exploits simple analytical formulae and hence is well-suited for
performing searches in large photometric databases.

\section{Ring transit geometry}
\label{sec:Geometry}

\begin{figure*}
{
\centering
\includegraphics [width=120mm] {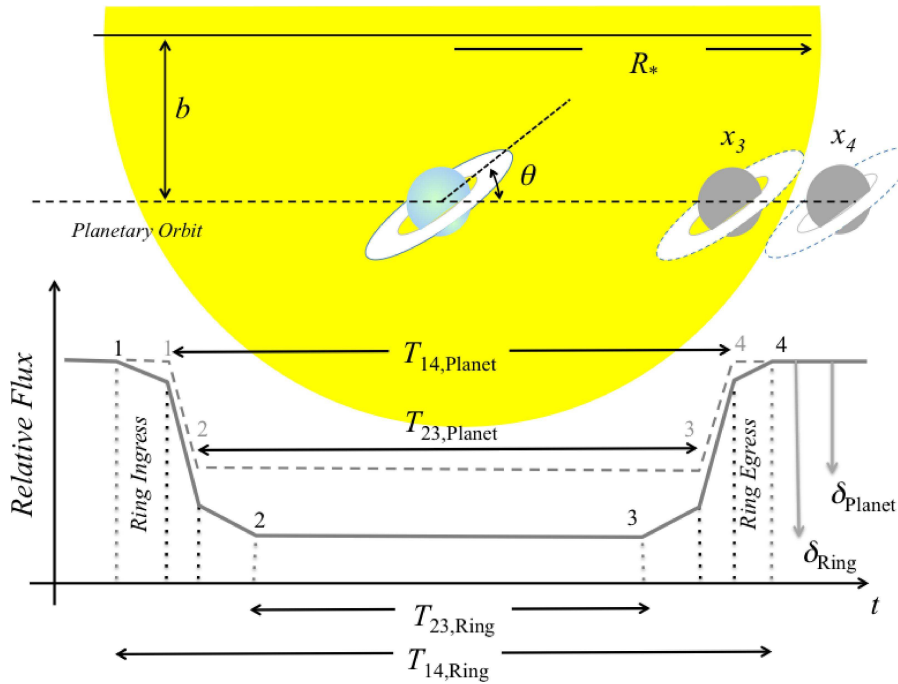}
\vspace{0.5cm}
\caption{Schematic ring and planetary transit {\hl geometry} and light
  curve. {\hl Sizes are not at scale. Transit depth ($\delta$) and
    total duration ($T_{14}$) are larger when an exoring is present,
    whereas the duration of full transit ($T_{23}$) is lower}.}
\label{fig:Geometry}
}
\end{figure*}

One of the most powerful properties of planetary transits is the
wealth of information they can provide by simply monitoring variations
in stellar brightness.

Assuming that the transiting object is spherical and other
conventional conditions, \citet{Seager03} showed that the planetary
radius $p\equiv \Rp/\Rs$, scaled orbital semimajor axis, $a/\Rs$, and
impact parameter, $b$, can be derived from three basic observables:

\begin{itemize}

\item[1.] transit depth, $\delta=(F_{\rm o}-F)/F_{\rm o}$ (where $F$
  and $F_{\rm o}$ represent the in-transit and out-of-transit stellar
  fluxes, respectively);

\item[2.] first-to-fourth contact transit duration, $\tT$;

\item[3.] second-to-third contact transit duration, $\tF$.

\end{itemize}

Figure \ref{fig:Geometry} schematically depicts the definition of
these quantities and the significant differences imposed by the
presence of planetary rings.

Our basic ring transit model relies on three basic assumptions.

\begin{itemize}

\item[1.] The planet and the star are spherical.

\item[2.] Rings are uniform and scatter/absorb light only between
  radii $\Rint=\fint\Rp$ and $\Rext=\fext\Rp$ (constant normal
  optical depth $\tau$).

\item[3.] Diffractive forward scattering in ring particles does not
  modify the basic transit parameters \citep{Barnes04}.

\end{itemize}

Under these assumptions, the transit depth of a ringed planet is given
by the ratio:

\beq{eq:p}
\delta=\frac{\ARp}{A_\star}=\frac{\ARp}{\pi\Rs^2},
\eeq

\noindent where $\ARp$ and $A_\star$ are the effective projected
ring-planet and stellar area.

The ring area is computed with the analytical expression (see Appendix
\ref{sec:AppAreas} for details)

\beq{eq:ARp}
\ARp = \pi R_p^2 + \pi [\reff^2(\fext) - \reff^2(\fint)] \,R_p^2,
\eeq

\noindent where $\reff$ is the {\it effective ring radius} and is
given by

\beq{eq:reff}
\reff^2(f)=\beta(\cos\iP)\times\;\left\{
\begin{array}{ll}
\ss f^2 \cos \iP - 1 & \ss f\cos \iP>1\\
\begin{array}{ll}
\ss f^2 \cos \iP & \frac{2}{\pi} {\ss\rm arcsin}\ss(y)+\\ \ss - &
\frac{2}{\pi} {\ss\rm arcsin}\ss(y f \cos\iP)
\end{array}
& \ss {\rm otherwise}
\end{array}
\right..
\eeq

Here, $\iP$ is the projected ring inclination ($\iP=90\dg$ if the ring
is edge on) and $y=\sqrt{f^2-1}/(f\sin\iP)$ is an auxiliary
variable. The term $\beta(\cos\iP)\equiv 1-e^{-\tau/\cos\iP}$ accounts
for the ring's effective absortion of stellar light \citep{Barnes04}.

To calculate ring transit durations, $\tT$ and $\tF$, we need to
compute the planet center horizontal coordinate, $x$, where the the
external ring or the planetary disk intersects the stellar limb at a
single point (see Figure \ref{fig:Geometry}).  Four $x$ values, the
contact positions, $x_i$ ($i=1,2,3,4$), fulfill this condition.

In the case of a non-ringed spherical planet, contact positions are
given by

\beq{eq:xP}
\left(\frac{x_{{\rm P},i}}{\Rs}\right)^2=(1\pm p)^2-b^2,
\eeq

\noindent {\hl where $\mhl{p=R_p/R_\star}$} and the ``+'' sign applies
to contacts 1 and 4 and the ``-'' sign to contacts 2 and 3.

If the contact point is on the edge of the external ring, then this
point satisfies a set of non-trivial algebraic/trigonometric
equations, whose solution is not expressable in a closed form (see
Appendix \ref{sec:AppIntersect}).  Combining several approximations,
we have found that in a wide range of ring and transit configurations,
the following analytical formula provides the value of $(x_i/\Rs)$
with a relative uncertainty no larger than a few percent:

\beq{eq:xRpm}
\left(\frac{x_{{\rm R},i}}{\Rs}\pm A\cos\theta\right)^2 \approx 
1-A^2(\sin\theta\mp b/A)^2(1-B^2/A).
\eeq

Here, $A=\fext p$ and $B=\fext p\cos\iP$ are the ring projected
semimajor and semiminor axes, and the signs now correspond to temporal
ordering.  The upper signs correspond to the ``leading'' contacts
(contacts 1 and 3) and the lower ones to the ``trailing'' contacts
(contacts 2 and 4).

The positive solution to both, Eqs. (\ref{eq:xP}) and (\ref{eq:xRpm}),
corresponds to contacts 3 and 4 and the negative solution to contacts
1 and 2.

Once the contact positions are calculated, the total duration of the
transit $\tT$ and the duration of full transit $\tF$ may be estimated
using \citep{Sackett99},

\beqn
\label{eq:tT}
\frac{a}{\Rs}\sin i\,\sin\left(\frac{2\pi}{P}\tT\right) & \approx &
\frac{x_4}{\Rs}-\frac{x_1}{\Rs}.\\
\label{eq:tF}
\frac{a}{\Rs}\sin i\,\sin\left(\frac{2\pi}{P}\tF\right) & \approx &
\frac{x_3}{\Rs}-\frac{x_2}{\Rs}.\\\nonumber
\eeqn

\noindent where $i$ and $P$ are the inclination and period of the
planetary orbit.

\section{Ring effect on observed planetary radius}
\label{sec:TransitDepth}

\begin{figure}
{ 
\centering 
\includegraphics[width=80mm]{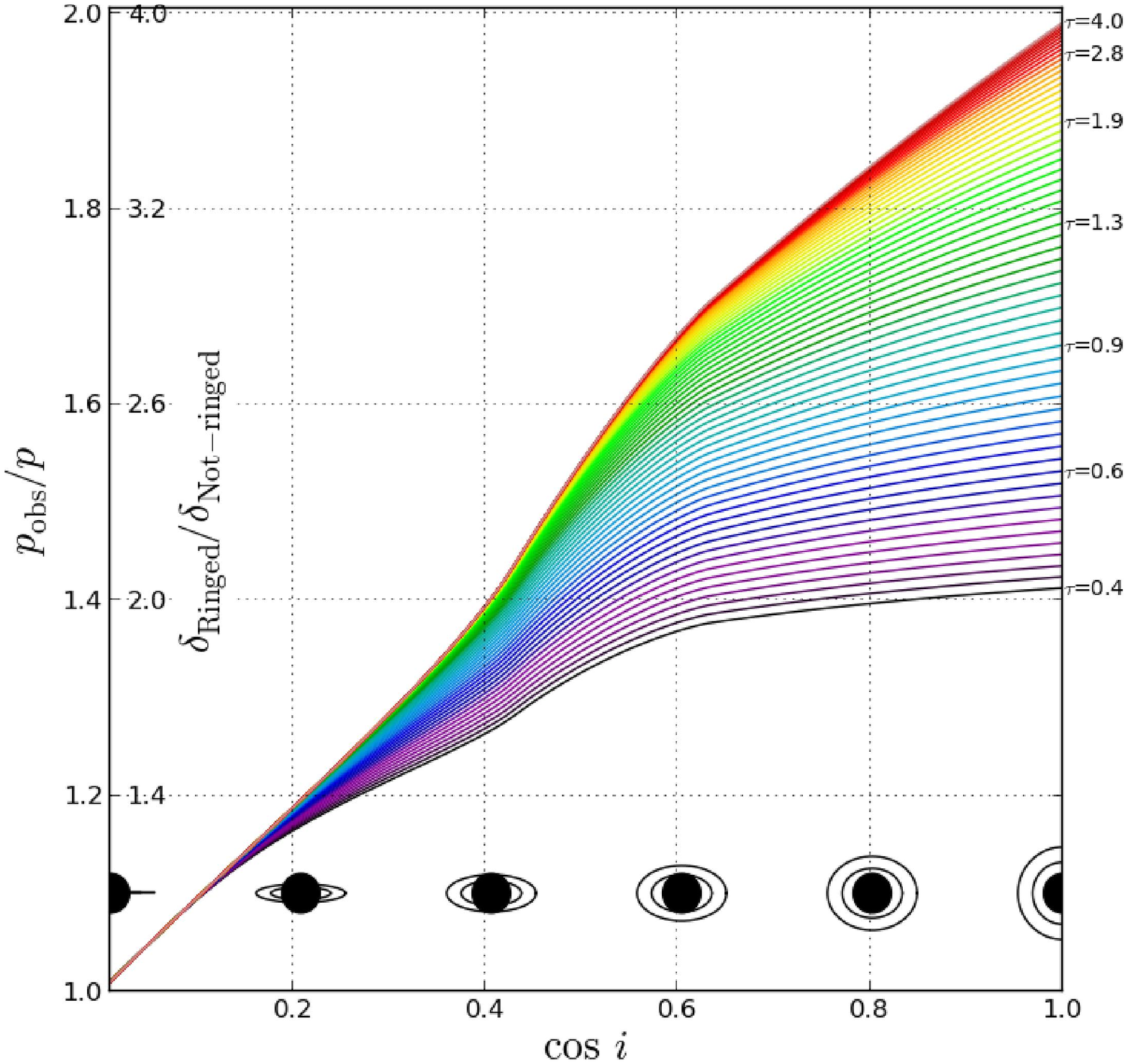}\\
\vspace{0.8cm}
\includegraphics[width=70mm]{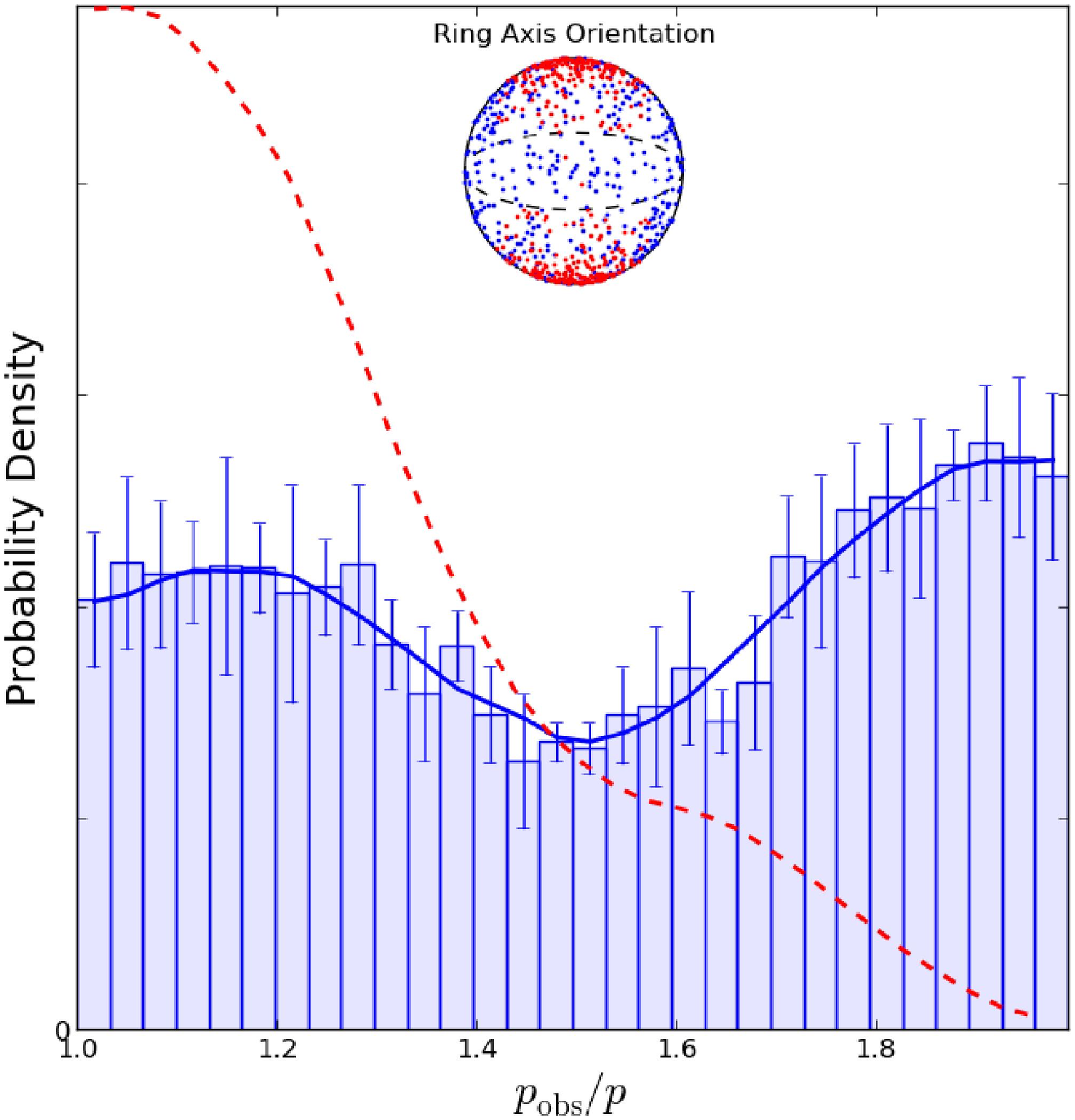}\vspace{0.2cm}
\caption{Upper panel: ratio of observed to true planetary radius {\hl
    as a function of ring projected inclination ($\cos\,i$), assuming
    different values for the ring normal opacity ($\tau$)}. Lower
  panel: probability distribution of this ratio {\hl assuming
    completely random ring orientations (blue histogram) and a more
    concentrated distribution of planetary obliquities (red dashed
    line). For illustration purposes, we show, in the inset diagram,
    the distribution of random axis orientations used to compute both
    distributions.  Each point represent the positions of the north and
    south pole of the planet for a given random orientation.}}
\label{fig:ObservedRadius}
}
\end{figure}

Ring transits could have a significant effect on the transit depth,
$\delta$.  To first order, $\delta$ provides the value of the apparent
or observed planetary radius, $\Rpobs$:

\beq{eq:Robs}
\Rpobs=\sqrt{\delta}.
\eeq

Thus, a significant overestimation of $\delta$ will also produce a
substantial overestimation of the observed radius.  In the absence of
any other evidence for the presence of exorings, an anomalously deep
transit may lead to the misclassification of a ringed planet as a
false-positive or yield a gross underestimation of the planetary
density.

To illustrate this, hereafter we assume the reference case of a planet
with the same radius as Saturn $\Rp=0.0836 R_\odot$ and rings of
similar size and properties, i.e., $\fint=1.58$ (inner edge of the
A-ring) and $\fext=2.35$ (outer edge of the B-ring).  We assume for
simplicity a value of $\tau=1$ (the opacity of B-ring ranges from
0.4-2.5, \citealt{Murray99}).  We also assume that the planet orbits a
solar-mass star in a circular orbit at $a=1$ AU.  The generalization
of these results to other planetary, ring, and orbital parameters is
trivial.

In Figure \ref{fig:ObservedRadius}, we plot the ratio of observed to
true planetary radius as obtained with Eqs. (\ref{eq:p}-\ref{eq:reff})
for our reference planet.

In the extreme case of face-on rings, Eqs. (\ref{eq:p}-\ref{eq:reff})
simplify to give

\beq{eq:RobsFO}
\lim_{\iR\to90\dg}\Rpobs=p\sqrt{1+(1-e^{-\tau})(\fext^2-\fint^2)}.
\eeq

The transits of a Saturn-like ringed planet are up to $\sim$3 times
deeper than these expected for a spherical non-ringed one. These deep
transits will be interpreted as being produced by a planet $\sim$1.7
times larger.  Additionally, if independent estimations of its mass
were also available, then the density of the planet would be
underestimated by a factor of $\sim$5.  Thus, instead of measuring
Saturn's density $\sim$0.7 g cm$^{-3}$, this planet would seem to have
an anomalously low density of $\sim$0.14 g cm$^{-3}$.  Even under more
realistic orientations ($\cos\iP\sim 0.2$), the observed radius will
be $\sim$20\% larger and the estimated density almost a half of the
real one.

To assess the effect of rings on the observed planetary radius for an
ensemble population of planets, we have calculated the probability
distribution of the ratio $(\Rpobs/p)$ assuming a uniform random ring
orientation (obliquity and azimuthal angle, as defined in
\citealt{Ohta09}).  The result is plotted in the lower panel of Figure
\ref{fig:ObservedRadius} (blue histogram).  The probability
distribution corresponding to predominantly low obliquities (following
a Fisher distribution in the case of a concentration parameter
$\kappa\sim 8$ and 2-$\sigma$ obliquity dispersion of $\sim$ 30$\dg$)
is also shown for comparison (red dashed line).

In the case of uniform random obliquities, more than 50\% of the
orientations lead to overestimations in planetary radius greater than
$\sim$50\%.  This implies that with no other clues for the existence
of rings around those planets, the bulk density of more than half of
them would be underestimated by up to a factor of 3.  With a more
concentrated distribution of obliquities, the resulting radius
distribution peaks at a null-effect of $(\Rpobs/p)\sim 1$ but still
exhibits considerable dispersion, with $\gtrsim$50\% of the cases
having observed radius anomalies $\gtrsim$ 30\% (density overestimated
by a factor $\gtrsim$2).

Deep transits are used as a criterion for flagging potential
false-positives in photometric surveys \citep{Batalha14,Burke14}.  If
a large unseen population of ringed planets exists, then we could have
many potential candidates buried among misclassified false-positives.
Therefore, we suggest that revisiting false-positive transits rejected
with this criterion could potentially lead to the discovery of the
first exoring population.

\section{The Photo-ring effect (PR-EFFECT)}
\label{sec:PR}

Under the simplified assumptions of the model presented in Section
\ref{sec:Geometry}, an analytic solution for the orbital semimajor
axis and impact parameter can be obtained from the basic transit
parameters \citep{Seager03}:

\beqn
\label{eq:a}
\left(\frac{a}{\Rs}\right)_{\rm obs} & \approx &
\frac{P}{2\pi}\frac{\delta^{1/4}}{(\tT^2-\tF^2)^{1/2}}\\
\label{eq:i}
b\sub{obs} & \approx &
\left[\frac{\tT^2(1-\delta)-\tF^2(1+\delta)}{\tT^2-\tF^2}\right]^{1/2},
\eeqn

where we also assume $\tT,\tF\ll P$, which is suitable for planets
with $a \gg \Rs$, i.e., those worlds with improved chances for
exorings.

Kepler's third law relates the orbital period, semimajor axis, and
stellar mass, such that we can calculate the mean stellar density:

\beq{eq:rhoobs}
\rhoobs=\frac{3\pi}{G}\frac{(a/\Rs)\sub{obs}^3}{P^2}
\eeq

In the case of a transiting spherical planet, the observed value of
$(a/\Rs)_{\rm obs}$, and hence $\rhoobs$, are accurate estimates of
their true values.  If the planet, however, has a ring, then $\delta$,
$\tT$ and $\tF$ will not be related by Equations (\ref{eq:a}) and
(\ref{eq:i}), and the observed quantities will differ from the true
ones.

A comparison of the observed density $\rhoobs$ with an independent
measurement, $\rhotrue$, such as that provided by stellar models,
asteroseismology \citep{Huber13}, or transits of other planetary
companions, could indicate the existence of exorings.

\begin{figure}
{ 
\centering 
\includegraphics[width=80mm]{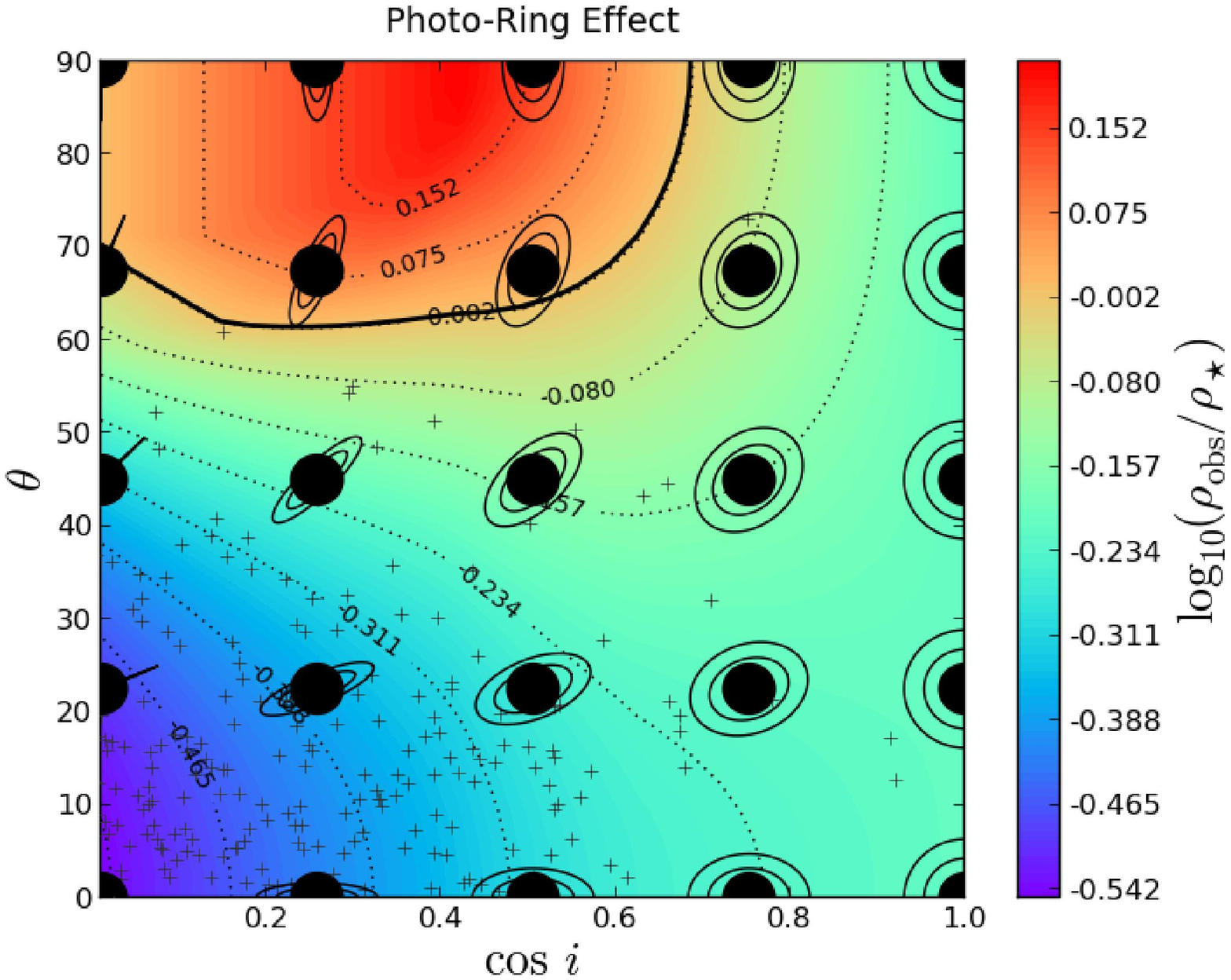}\\
\vspace{0.5cm}
\includegraphics[width=70mm]{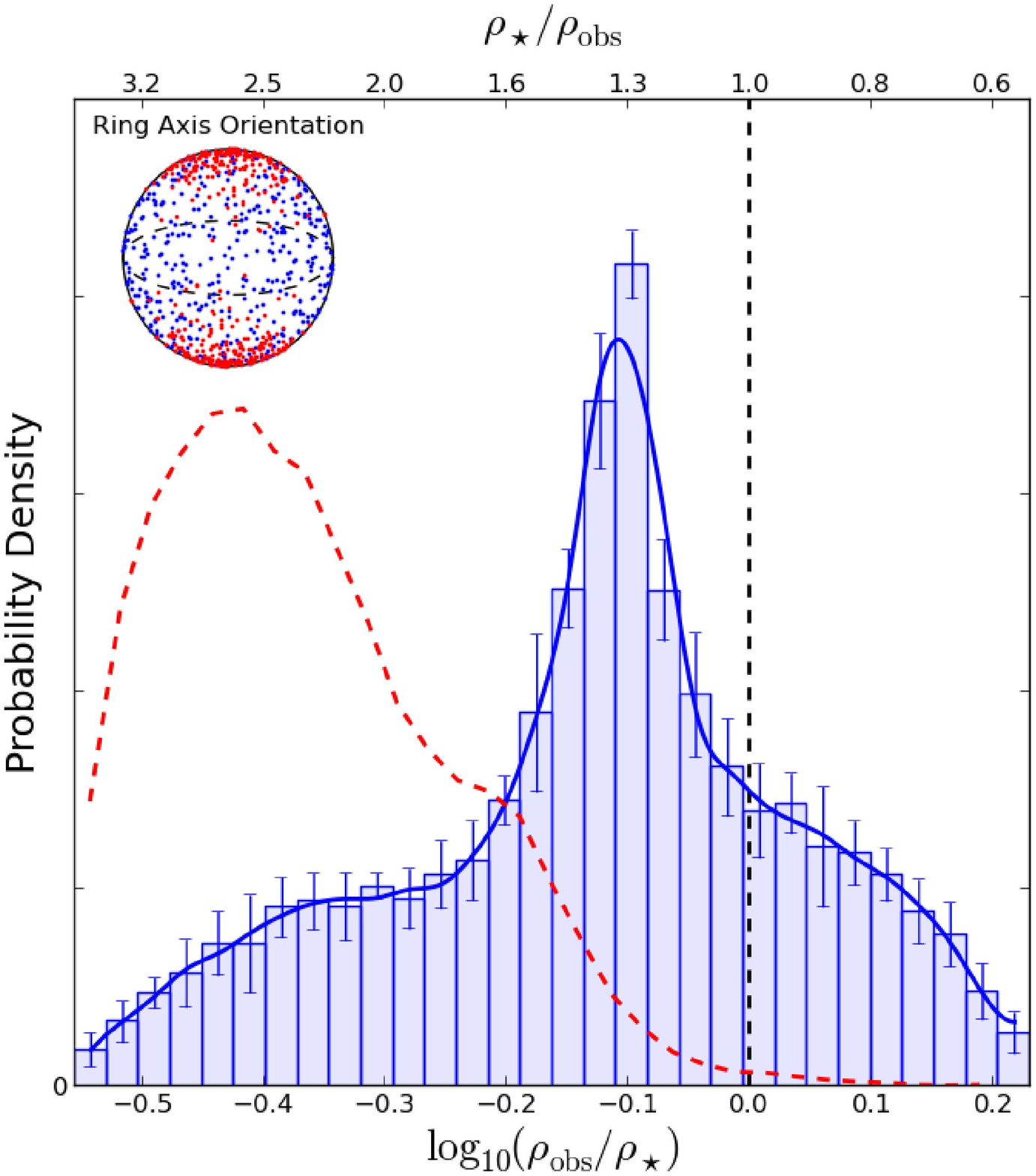}
\vspace{0.2cm}
\caption{Upper panel: contour plot of the PR effect. {\hl The crosses
    correspond to actual values of the projected inclinations for a
    subsample drawn from a low obliquity population (red dots in the
    inset plot of the lower panel)}.  Lower panel: probability
  distribution of the PR effect {\hl for completely random
    orientations (blue) and for a more concentrated distribution of
    obliquities (red)}.}
\label{fig:PhotoRing}
}
\end{figure}

In the upper panel of Figure \ref{fig:PhotoRing}, we show contours in
the plane of the projected ring orientation of this so-called
PR-effect.

Since $\tT$ ($\tF$) is larger (smaller) when a ring is present (see
Figure \ref{fig:Geometry}), the observed semimajor axis $\aobs$ could
be over- or underestimated. For large tilts ($\theta\sim 90\dg$) and
inclinations ($\cos\iR\sim 0$), the transit duration will be the same
as that of a non-ringed planet.  Consequently, the increased transit
depth $\delta$ will be the dominant effect in Eq. (\ref{eq:a}) and the
observed density will be overestimated (upper left region in Figure
\ref{fig:PhotoRing}).

For most values of the ring's projected inclination and tilt, however,
the transit duration will be modified to a larger extent than the
depth, causing the semimajor axis and density to be underestimated.
Therefore, rings tend to produce a negative PR effect (underestimation
of stellar density) and this could be used to distinguish it from
other asterodensity profiling (AP) effects discussed in
\citet{Kipping14}.

We can use Eqs. (\ref{eq:xP})-(\ref{eq:tF}) to calculate an analytical
expression for the maximum PR effect expected for a given external
disk radius and impact parameter:

\beq{eq:PRmax}
\left(\frac{\rhoobs}{\rhotrue}\right)_{\rm max}\approx
\fext^{-3/2}(1-b^2)^{-3/4}
\eeq

For our reference case, $\fext=2.35$, $b=0$, and the maximum value of the
PR effect is $\log_{10}(\rhoobs/\rhotrue)\sim -0.6$.  This corresponds
to an underestimation of the stellar density by a noticeable factor of
$\sim 4$.

We have calculated the probability distribution of the PR effect
assuming similar priors to those used earlier.  The results are shown
in the lower panel of Figure \ref{fig:PhotoRing}.

In the uniform case, the $\PR$ distribution is strongly peaked around
a negative value, $\log_{10}(\rhoobs/\rhotrue)\sim -0.2$.  In
contrast, if we consider obliquities no larger than $\sim$30$\dg$,
then the peak shifts to $\log(\rhoobs/\rhotrue)\sim -0.45$,
corresponding to a notable difference by a factor of $\sim$ 3
between the observed and true density.

The only other AP effects which can cause comparably large deviations
are the photo-eccentric (PE) and the photo-blend effects, for which
the former tends to cause a positive AP deviation and the latter a
negative \citep{Kipping14}. In the case where blended companions can
be excluded (e.g., through high-resolution imaging), then the ensemble
distribution of the PR-effect should therefore be distinguishable from
the other AP effects.

\section{Discussion and Conclusions}
\label{sec:Discussion}

In this Letter, we have presented a novel method for identifying
exorings.  Our method does not require a complex fit of transit light
curves, instead relying on simple, analytic, and computationally
efficient numerical procedures.

Our technique exploits the substantial impact that rings produce on
the transit depth and duration, effects that can be independently
confirmed by the measurement or estimation of other relevant
astrophysical properties (e.g., stellar and planetary density).

Interestingly, the two effects we seek (anomalous transit depths and
PR-effect) are {\hl complementary} with respect to the orientation of
the ring plane.  For large inclinations and obliquities (face-on
rings), the effect on transit depths is significant whilst the
photo-ring effect is negligible.  Alternatively, if rings {\hl have}
relatively low obliquities (egde-on rings), then the PR-effect will be
considerable but the depth anomaly small.

Accordingly, three basic {\hl complementary} strategies are proposed
to identify exorings among the confirmed transiting exoplanets and
candidates.

\begin{enumerate}

\item Search the confirmed transiting planets with anomalously low
  densities.

\item Search transiting objects that have been tagged as
  false positives due to anomalously large transit depths.

\item Search transit signals for which a negative AP effect is
  observed.

\end{enumerate}

Besides aiding observers seeking interesting individual systems, the
effects described here are well-suited for inferring a population of
ringed planets. Ensemble studies using AP are in their infancy, but
\citet{Sliski14} recently conducted the first such analysis on a
sample of 41 single Kepler planetary candidates with asteroseismically
constrained host stars. In this limited sample, largely focused on
short-period planets, the authors conclude that the 31 dwarf stars
yield a broad AP distribution about zero, consistent with the expected
photo-eccentric variations, whereas as the objects associated with
giant stars are likely orbiting different stars altogether. A larger
sample, including longer-period planets, would provide the opportunity
to seek the expected offset due to the PR-effect too. This would
require a reliable independent measure of the stellar density for
stars too faint for asteroseismology, perhaps using methods such as
``flicker'' \citep{Kipping14c}.

Low-density planets are also interesting targets when seeking
exoplanetary rings. Over 10\% of the already confirmed planets have
estimated
densities\footnote{\href{http://exoplanet.eu}{http://exoplanet.eu}}.
Most of them are planets larger than Neptune, and a non-negligible
fraction have anomalously low densities below $\sim$0.3 g cm$^{-3}$.
Although several successful explanations have been devised for
reconciling observed low densities with planetary interior and thermal
evolutionary models \citep{Miller09}, other anomalies still remain and
may be worth a further analysis along the lines suggested here.

With the exception of {\it Kepler}-421b \citep{Kipping14b}, all of the
confirmed planets and most of Kepler candidates are inside the
so-called snow line.  Although icy rings, such as those observed
around Saturn and Solar System giant planets, seem unlikely inside
this limit, the existence of ``warm'' rocky rings at distances as
short as $\sim$0.1 AU, are not dynamically excluded
\citep{Schlichting11}.

We stress that the method presented here is {\hl complementary} to the
methods developed to discover exorings through detailed light curve
modeling \citep{Barnes04,Ohta09,Tusnski11}.  As explained earlier, the
role of these methods will be very important once a suitable list of
potential exoring candidates is created.  It is, however, also
important to note the great value of light curve models developed
under the guiding principle of computational efficiency
(semi-analytical formulae, efficient numerical procedures, etc), such
as the basic models presented here.

Our simple technique is suitable for surveying entire catalogs of
transiting planet candidates for exoring candidates, providing a
subset of objects worthy of more detailed light curve
analysis. Moreover, the technique is highly suited for uncovering
evidence of a population of ringed planets by comparing the radius
anomaly and PR-effects in ensemble studies.  To aid the
community, we provide the publicly available code at
\href{http://github.org/facom/exorings}{http://github.org/facom/exorings}
to simulate the novel effects described in this work.

\section*{Acknowledgements}

J.I.Z. is supported by Vicerrectoria de Docencia, Universidad de
Antioquia, Estrategia de Sostenibilidad 2014-2015 de la Universidad de
Antioquia and by the Fulbright Commission, Colombia.  J.I.Z. thanks
the Harvard-Smithsonian Center for Astrophysics for its hospitality
while this work was carried out.  D.M.K. is supported by the Menzel
fellowship.  M.S. is supported by CODI/UdeA and J.A.A. by the Young
Researchers program of the Vicerrectoria de Investigacion/UdeA.  We
thank to our refere, Jason Barnes, for his comments and useful
suggestions.

\appendix

\section{Area of the Ring}
\label{sec:AppAreas}

\begin{figure*}
{
\centering
\includegraphics [width=80mm] {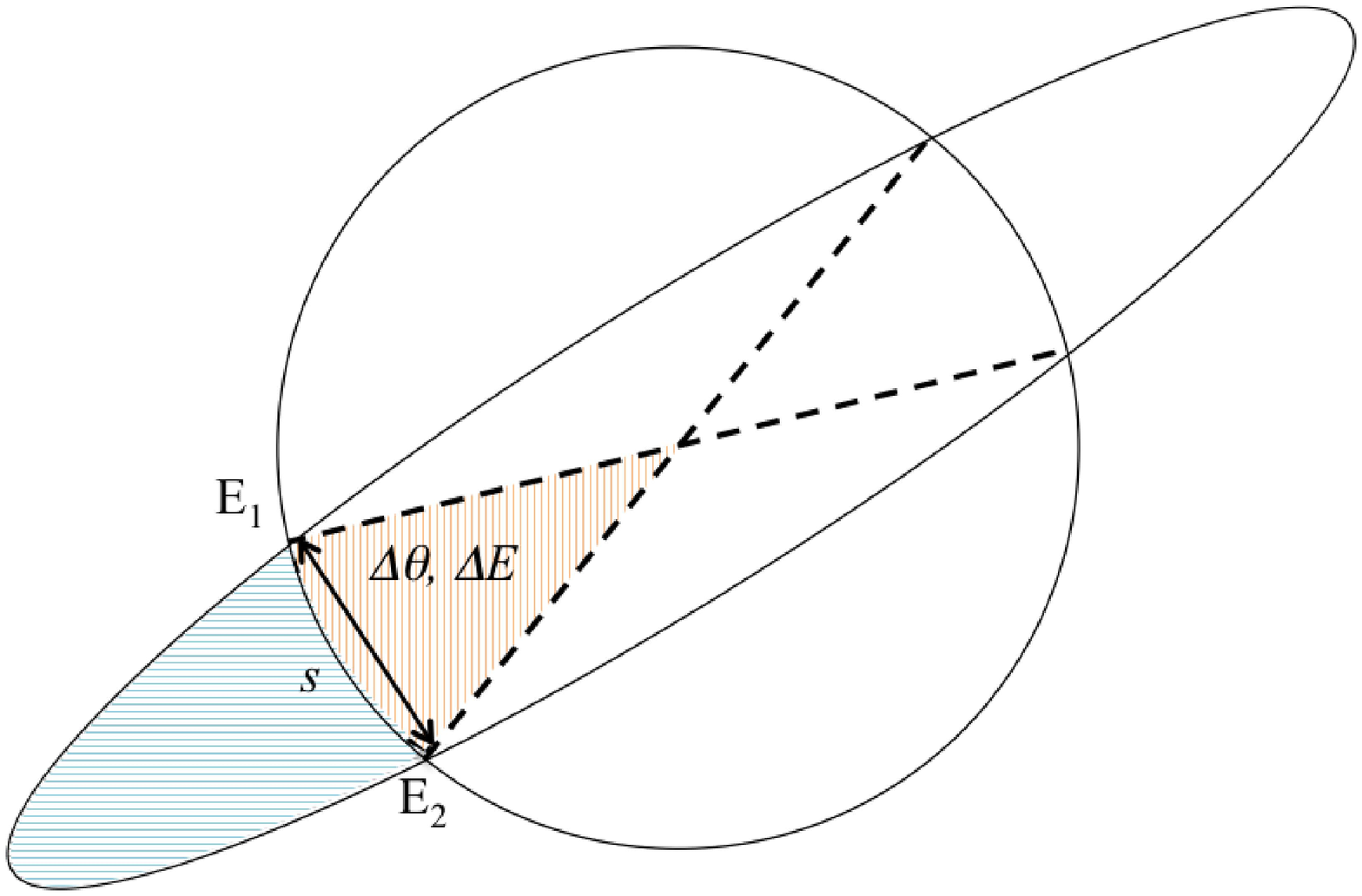}
\hspace{0.5cm}
\includegraphics [width=80mm] {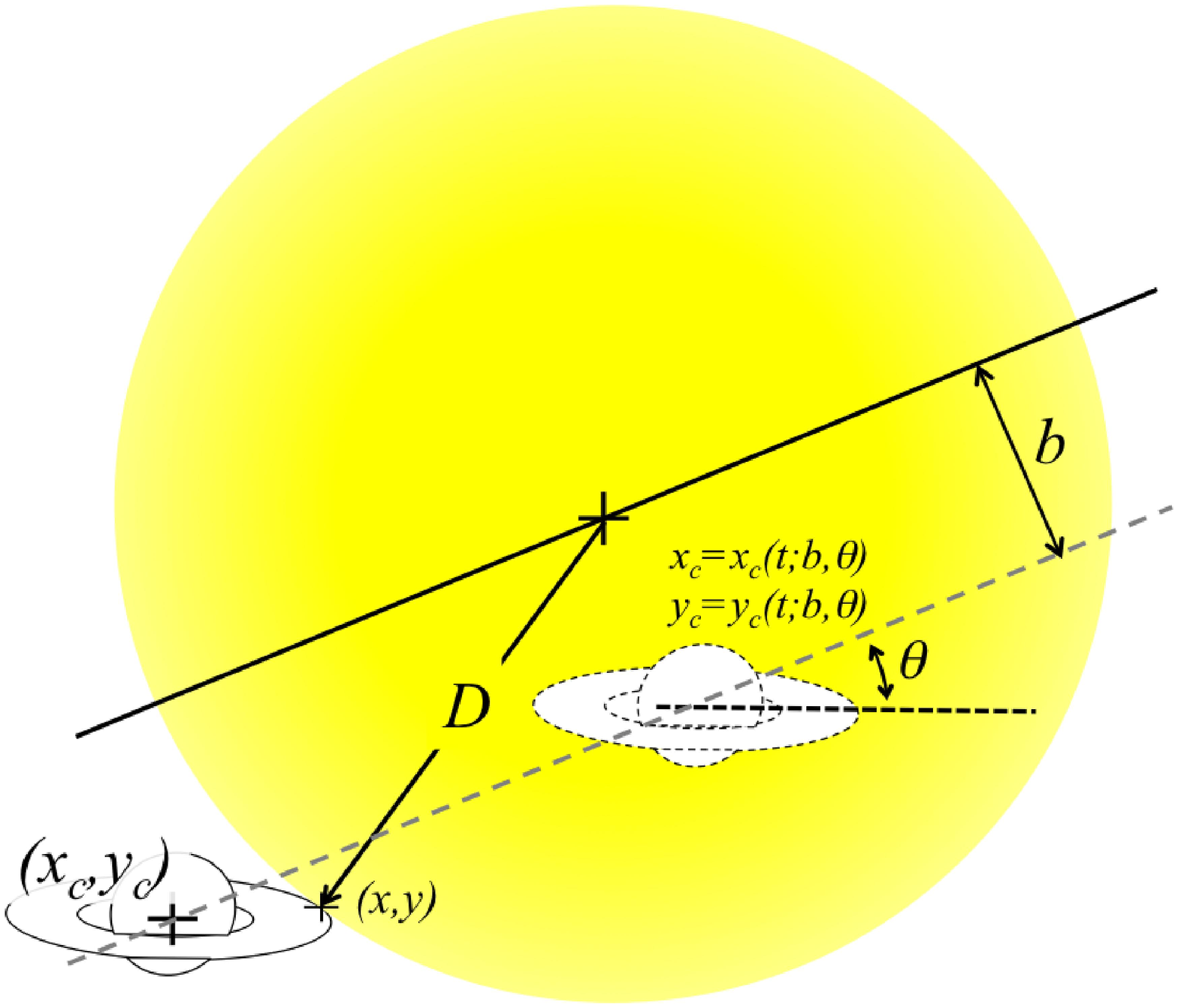}
\vspace{1.0cm}
\caption{{\hl Geometrical} constructions used to compute ring area
  (left panel) and contact position (right panel).  {\hl For contact
    position, a rotated reference system where the rings are
    horizontal considerably eases the analytic
    derivations}.\vspace{0.2cm}}
\label{fig:Construction}
}
\end{figure*}

The area obscured only by the ring is computed substracting the areas
of the ellipse and circle sectors limited by interesection points
E$_1$ and E$_2$ (Figure \ref{fig:Construction}):

\beq{eq:RingArea}
A_{\rm ring}=2(A_{\rm e}-A_{\rm c}).
\eeq

The area of the circle and ellipse sectors follow from the
Cavallieri's Principle\footnote{A. Bogomolny, Interactive Mathematics
  Miscellany and Puzzles
  \href{http://www.cut-the-knot.org/Generalization/Cavalieri2.shtml}{http://www.cut-the-knot.org/Generalization/Cavalieri2.shtml},
  Accessed 2015 January 21}:

\begin{eqnarray}
A_{\rm c} & = & \frac{1}{2}\Delta\theta\;R^2\\\nonumber
A_{\rm e} & = & \frac{1}{2}\Delta E\;A B,
\label{eq:Sectors}
\end{eqnarray}

\noindent where $A$ and $B$ are the apparent semimajor and semiminor
axes of the ring, $\Delta\theta$ is the angle subtended by the sector,
and $\Delta E$ is the eccentric anomaly difference between E$_1$ and
E$_2$.

The length of the segment joining E$_1$E$_2$ is\footnote{Weisstein, Eric
  W. MathWorld. \href{http://mathworld.wolfram.com/Circle-EllipseIntersection.html}{http://mathworld.wolfram.com/Circle-EllipseIntersection.html}. ,
  Accessed 2015 January 21}:

\beq{eq:s}
\frac{s}{2}=B \sqrt{\frac{A^2-R^2}{A^2-B^2}}
\eeq

Simple trigonometrical relationships and the parametric equation of
the ellipse provide us with the expressions for $\Delta\theta$ and
$\Delta E$:

\beqn
\Delta\theta & = & 2\arcsin\left(\frac{s}{2R}\right)\\\nonumber
\Delta E & = & 2\arcsin\left(\frac{s}{2B}\right).
\label{eq:deltaE}
\eeqn

Inserting Equations (\ref{eq:deltaE}) into \ref{eq:Sectors}, the areas
are finally given by

\beqn
A_{\rm c} & = & R^2
\arcsin\left(\cos\iP\frac{\sqrt{f^2-1}}{f\sin\iP}\right)\\\nonumber
A_{\rm e} & = & f^2 \cos\iP\;R^2
\arcsin\left(\frac{\sqrt{f^2-1}}{f\sin\iP}\right).\\\nonumber
\eeqn

From these areas and Eq. (\ref{eq:RingArea}), the effective ring
radius in Eq. (\ref{eq:reff}) follows trivially.

\section{Ring contact positions}
\label{sec:AppIntersect}

Points over the external ring in the right panel of Figure
\ref{fig:Construction} obey the following parametric equations:

\beqn
x(E,t) & = & A\cos E+x_c(t)\\\nonumber
y(E,t) & = & B\sin E+y_c(t)
\eeqn 

\noindent where $E$ is the eccentric anomaly, $x_c(t),y_c(t)$ are the
instantaneous coordinates of the planet center, and $t$ is an arbitrary
parameter (time for instance).

Contact positions are those for which the distance $D$ to origin,

\beq{eq:D2}
D^2(E,t)=[A\cos E+x_c(t)]^2+[B\sin E+y_c(t)]^2,
\eeq

\noindent obeys two conditions:

\beqn
\label{eq:Et}
\left.\frac{\partial D^2}{\partial E}\right|_{E_i,t_i} & = & 0\\\nonumber
D^2(E_i,t_i) & = & \Rs^2
\eeqn

\noindent or explicitly:

\beqn
\label{eq:Eiti}
A^2\cos^2 E_i + B^2\sin^2E_i + x_c^2(t_i) + y_c^2(t_i) + 2 [A\,x_c(t)
  cos E_i + B\, y_c(t_i) \sin E] & = & 0\\\nonumber
(B^2-A^2)\sin E_i\cos E_i-A\,x_c(t_i) \sin E + B\,y_c(t_i) \cos E_i - \Rs^2 & = & 0.
\eeqn

The solutions to these trigonometric equations, altough {\hl
  programmable}, are not expressable in closed form.

The ``empirical'' formula in Eq. (\ref{eq:xRpm}) was obtained after
the approximations $A\gg B$ and $|\cos E_i|\sim 1$.  {\hl This only
  works for} $\cos\iP\sim 0$ and $\theta\sim 0$.  We have verified,
however, that contact times estimated with Eq. (\ref{eq:xRpm}),
(\ref{eq:tT}) and (\ref{eq:tF}) are off by $\lesssim$1\% in the case
of contacts 3 and 4 for most ring projected orientations (provided
$\theta>0$) and $\lesssim$10\% for contacts 1 and 2.  {\hl In the
  cases when $\theta\gtrsim 60\dg$ or $\cos\,i\gtrsim 0.5$ a numerical
  solution for Eqs. (\ref{eq:Eiti}) is required to attain an
  acceptable precision.}



\end{document}